\documentclass[sigconf]{acmart}

\usepackage{balance}
\usepackage{subcaption}
\usepackage{multirow}
\usepackage{enumitem}
\usepackage{listings}

\usepackage{tikz}

\usetikzlibrary{calc}
\newcommand{\tikzmark}[1]{\tikz[overlay,remember picture] \node (#1) {};}
\newcommand*{\DrawArrow}[3][]{%
    \begin{tikzpicture}[overlay,remember picture]
        \draw [-latex, #1] ($(#2)+(0.1em,0.5ex)$) to ($(#3)+(0,0.5ex)$);
    \end{tikzpicture}%
}%

\theoremstyle{definition}

\usepackage{multicol}
\usepackage{url}

\AtBeginDocument{%
  }

\copyrightyear{2025}
\acmYear{2025}
\setcopyright{rightsretained}
\acmConference[FPGA '25]{Proceedings of the 2025 ACM/SIGDA International Symposium on Field Programmable Gate Arrays}{February 27--March 1, 2025}{Monterey, CA, USA}
\acmBooktitle{Proceedings of the 2025 ACM/SIGDA International Symposium on Field Programmable Gate Arrays (FPGA '25), February 27--March 1, 2025, Monterey, CA, USA}
\acmDOI{10.1145/3706628.3708871}
\acmISBN{979-8-4007-1396-5/25/02}

\makeatletter
\gdef\@copyrightpermission{
  \begin{minipage}{0.2\columnwidth}
   \href{https://creativecommons.org/licenses/by/4.0/}{\includegraphics[width=0.90\textwidth]{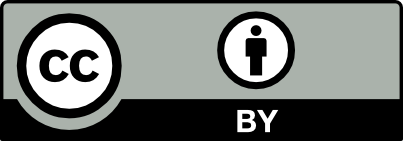}}
  \end{minipage}\hfill
  \begin{minipage}{0.8\columnwidth}
   \href{https://creativecommons.org/licenses/by/4.0/}{This work is licensed under a Creative Commons Attribution International 4.0 License.}
  \end{minipage}
  \vspace{5pt}
}
\makeatother

\begin{document}

\title{Dynamic Loop Fusion in High-Level Synthesis}

\author{Robert Szafarczyk}
\affiliation{%
  \institution{University of Glasgow}
  \city{Glasgow}
  \country{UK}
}
\email{robert.szafarczyk@glasgow.ac.uk}

\author{Syed Waqar Nabi}
\affiliation{%
  \institution{University of Glasgow}
  \city{Glasgow}
  \country{UK}
}
\email{syed.nabi@glasgow.ac.uk}

\author{Wim Vanderbauwhede}
\affiliation{%
  \institution{University of Glasgow}
  \city{Glasgow}
  \country{UK}
}
\email{wim.vanderbauwhede@glasgow.ac.uk}


\begin{abstract}

Dynamic High-Level Synthesis (HLS) uses additional hardware to perform memory disambiguation at runtime, increasing loop throughput in irregular codes compared to static HLS.
However, most irregular codes consist of multiple sibling loops, which currently have to be executed sequentially by all HLS tools.
Static HLS performs loop fusion only on regular codes, while dynamic HLS relies on loops with dependencies to run to completion before the next loop starts.

We present dynamic loop fusion for HLS, a compiler/hardware co-design approach that enables multiple loops to run in parallel, even if they contain unpredictable memory dependencies.
Our only requirement is that memory addresses are monotonically non-decreasing in inner loops.
We present a novel program-order schedule for HLS, inspired by polyhedral compilers, that together with our address monotonicity analysis enables dynamic memory disambiguation that does not require searching of address histories and sequential loop execution.
Our evaluation shows an average speedup of 14$\times$ over static and 4$\times$ over dynamic HLS.
\end{abstract}




\maketitle


\section{Introduction} \label{sec:introduction}
\begin{figure}[t]
\centering
 \begin{subfigure}[b]{0.45\textwidth}
     \centering
     \begin{verbatim}
    for (i = 0; i < N; ++i) 
        A[f(i)] = workA(A[f(i)]);
    for (j = 0; j < M; ++j) 
        B[g(j)] = workB(A[g(j)]);
\end{verbatim}
     \caption{Two loops with non-affine access patterns.}
     \label{fig:Pipeline_a}
 \end{subfigure}
 \hfill
 \begin{subfigure}[b]{0.45\textwidth}
     \centering
     \includegraphics[width=\textwidth]{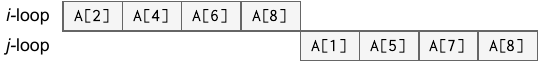}
     \caption{Pipeline achieved by current static and dynamic HLS tools.}
     \label{fig:Pipeline_b}
 \end{subfigure}
 \hfill
 \begin{subfigure}[b]{0.45\textwidth}
     \centering
     \includegraphics[width=\textwidth]{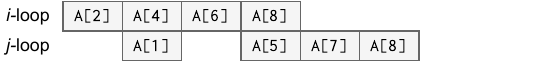}
     \caption{Pipeline achieved by our work.}
     \label{fig:Pipeline_c}
 \end{subfigure}

\caption{Dynamic Loop Fusion enables fine-grained parallelism \textit{across} loops with memory dependencies.}
\Description[Three sub-figures. Sub-figure a shows motivating C code. Sub-figures b shows the pipeline generated by current tools. Sub-figures c shows the pipeline generated by our work.]{Three sub-figures. Sub-figure a shows motivating C code. Sub-figures b shows the pipeline generated by current tools. Sub-figures c shows the pipeline generated by our work.}
\label{fig:Pipeline}
\end{figure}

High-Level Synthesis (HLS) increases designer productivity, makes code more maintainable, accelerates verification, and makes design space exploration easier \cite{google_vcu}.
However, this is usually only true for regular codes where the compiler can discover instruction- and memory-level parallelism statically \cite{modulo_sched, modulo_sched_canis}. 
Domains like graph analytics and sparse linear algebra contain irregular codes with unpredictable memory dependencies and control flow, which break the traditional static scheduling approach.
This prompted research into dynamically scheduled HLS \cite{josipovic_dynamatic_2022} and approaches to combine it with existing industry-grade static HLS compilers \cite{dass, szafarczyk_fpl23}.

Dynamic HLS uses load-store queues (LSQs) to perform memory disambiguation at runtime \cite{lsq_2014, Josipovic_Brisk_Ienne_2017, szafa_fpt23, unified_memory_dependency_spec_hls, adaptive_loop_pipelining, dai_speculative_data_hazards}.
These works effectively pipeline single loops with arbitrary memory dependencies, but they have to sequentialize multiple loops if they share a memory dependency.
For example, they would sequentialize the \textit{i}- and \textit{j}-loops in figure~\ref{fig:Pipeline_a}, resulting in the figure~\ref{fig:Pipeline_b} pipeline.
But, as shown in figure~\ref{fig:Pipeline_c}, there might be plenty of parallelism \textit{across} the two loops.

There are two reasons why current dynamic HLS tools have to sequentialize these loops.
Firstly, they use a program-order schedule that relies on loops to run to completion before the next loop starts.
For example, the LSQ used in Dynamatic HLS sequentializes LSQ requests based on the program order of basic blocks \cite{Josipovic_Brisk_Ienne_2017}; other approaches carry explicit dependencies through the pipeline, preventing downstream loops from starting without resolving the dependency \cite{straight_to_the_queue, szafa_fpt23}.
Secondly, they rely on the checking of address histories to detect hazards, without making any assumptions about the underlying address distributions.
This makes them general, but requires them to wait for all addresses from one loop to be produced before they can start processing the next loop.
These are two \textit{key challenges} that we tackle in this paper. 

Static loop fusion 
also fails to fuse the loops in our figure~\ref{fig:Pipeline_a} example, because the fused loop may introduce a negative dependency distance \cite{Optimizing_Compilers}---the compiler gives up if it cannot prove that $f(i) = g(j) \implies i < j$.
This is assuming that the $f(i)$ and $g(i)$ functions can be analyzed by the compiler in the first place.
If that is not the case, e.g., if they involve an array access, then loop fusion is also not applied.


Our dynamic loop fusion approach can automatically synthesize a Read After Write (RAW) check that will protect the $A[g(j)]$ read in the figure~\ref{fig:Pipeline} code, achieving the fine-grained inter-loop parallelism from figure~\ref{fig:Pipeline_c}.
We decouple each loop into an independently scheduled Processing Element (PE).
Memory dependencies across loops are handled in a Data Unit (DU) specialized by our compiler for the program.
Our only requirement is that the \texttt{f(i)} and \texttt{g(j)} functions are monotonically non-decreasing in the innermost loop (outer loops can be non-monotonic).
This is a weaker requirement than the affine functions expected by static loop fusion, allowing us to fuse more loops, including codes with data-dependent addresses. 

To the best of our knowledge, we are the first to propose dynamic memory disambiguation that can work \textit{across} loops.
We make the following contributions:
\begin{itemize}
    \item A compiler pass to decouple loop nests into PEs. A PE is further decoupled into an Address Generation Unit (AGU) and a Compute Unit (DU) following the decoupled access/execute (DAE) architecture (section \ref{sec:baseline_arch}).
    \item A compiler analysis, based on the chain of recurrences theory \cite{scev_bachmann, scev_engelen}, that checks if addresses are monotonically non-decreasing in inner loops, and that detects non-monotonic outer loops (section \ref{sec:address_monotonicity}). 
    \item A hardware-efficient program-order schedule representation that does not require sequentializing loops. We show how the compiler instruments AGUs with instructions that generate the schedule for each memory operation. We also show how non-monotonic outer loops can be integrated with our schedule (section \ref{sec:schedule}). 
    \item A parameterizable DU performing dynamic memory disambiguation across loops. We show how the compiler can specialize the DU given the dependency graph of the program and the address monotonicity analysis. We discuss how the DU optimizes DRAM bandwidth by using dynamic coalescing and on-chip store-to-load forwarding (section \ref{sec:dataUnit}).
    \item An evaluation on irregular applications showing an average speedup of 14$\times$ over static HLS and 4$\times$ over dynamic HLS. We discuss which codes benefit from dynamic fusion and we study the impact of store-to-load forwarding (section \ref{sec:Evaluation}).
\end{itemize}

\section{Background}

In this paper, we focus on codes using DRAM, as its unpredictable latency and limited bandwidth pose greater challenges than BRAM.
There is no fundamental reason why we could not protect BRAM or use a memory hierarchy with BRAM caches, which we briefly discuss in section \ref{sec:limit_and_future}.

In this section, we describe FPGA streaming architectures commonly used with DRAM.
We discuss techniques to optimize DRAM bandwidth in irregular codes that inform the design of our DU.
And we describe existing loop fusion approaches and their compiler theory, informing the design of our program-order schedule representation.

\subsection{Baseline Streaming Architecture} \label{sec:baseline_arch}
\begin{figure}[t]
\centering
\includegraphics[width=0.475\textwidth]{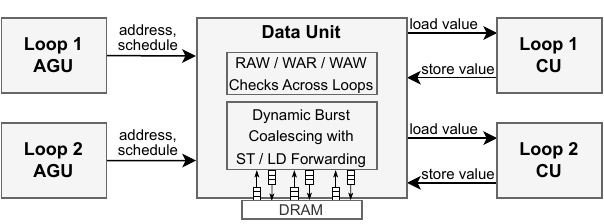}

\caption{An example DAE streaming FPGA architecture.}
\label{fig:ArchitectureTemplate}
\end{figure}


Streaming FPGA architectures are a popular choice for implementing DRAM-based codes \cite{tapa, fleet, st_accel, stencil-flow, tytra_ijpp}.
They decouple memory accesses and compute into separate PEs, either automatically \cite{stencil-flow, tytra_ijpp} or manually \cite{tapa, fleet, st_accel}.
The use of a streaming architecture is predicated on an accurate memory dependency analysis so that memory shared between PEs can be transformed into FIFO communication.
If the analysis fails, as it invariably does for irregular codes, then the shared data has to be communicated via DRAM and the execution of PEs has to be sequentialized, thus losing much of the benefits of using a streaming FPGA architecture.

To tackle the problem of irregular memory accesses, we propose to use a compiler-parameterized DU, shown in figure~\ref{fig:ArchitectureTemplate}, that protects memory shared across loops by performing dynamic memory disambiguation at runtime.
The DU interfaces with DRAM, but is also able to directly forward values from producer to consumer PEs if the respective load/store operations exhibit temporal locality, thus saving DRAM bandwidth as in traditional streaming FPGA architectures.

\subsubsection{Using DRAM Bandwidth Efficiently}
We use Altera's DRAM IP generated by its HLS compiler to implement DRAM load/store units (LSUs).
Our DU can have multiple LSUs connected to the DRAM controller using a ring topology, depending on the number of load/store operations in the input program.
To use DRAM bandwidth efficiently, the LSUs coalesce multiple loads/stores into one wide request to the memory controller in order to use the full DDR channel width (512-bit in our case).
To achieve this for codes with irregular access patterns, the LSUs use additional logic and buffering to perform coalescing dynamically \cite{dynamic_burst_lsu_intel, request_coalesce_serve}.
DRAM requests are buffered until the largest possible burst can be made.
If no new requests arrive in $N$ consecutive cycles, then an incomplete burst is made (in our case $N=16$).

Asynchronous address supply is essential for efficient use of DRAM, because of the high access latency, and to allow the dynamically bursting LSU to look ahead in the address stream.
Streaming FPGA architectures achieve this by following the decades-old DAE principle \cite{decoupled_access_exec}, where the address generation is decoupled into its own thread of execution, running ahead of the compute threads that consume and produce values \cite{fifer, decoupled_memory_prefetching, decoupled_memory_wawrzynek, HLS_runahead}.
Note that dynamic loop fusion is not limited to DAE architectures; it can be realized in other model of computations, e.g., with dynamic dataflow \cite{josipovic_dynamatic_2022}.

\begin{figure}[t]
\centering
     \includegraphics[width=0.43\textwidth]{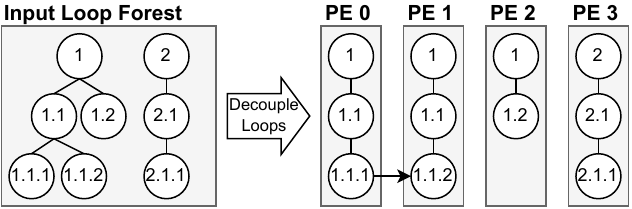}
     \caption{Example decoupling of a loop forest. A leaf loop is decoupled into its own PE, which includes loop control of the outer loops. Parent loop body instructions are included only if they come before the leaf loop in the topological order. FIFOs are used to communicate scalar data dependencies (e.g. from loop 1.1.1 in PE 0 to loop 1.1.2 in PE 1). FIFOs are written in the loop exit block and read in the loop pre-header block. Each loop PE might have an AGU producing memory requests fo the accesses in that loop PE.}
     \label{fig:LoopDecoupling}
\end{figure}

\subsubsection{DAE Transformation} 
A DAE architecture is automatically generated by our compiler.
Given a forest of loop trees, loop CUs and AGUs are decoupled into their own PEs following the strategy from figure~\ref{fig:LoopDecoupling}.
AGUs feed addresses to the DU; the DU sends load values to and receives store values from CUs.
All communication is FIFO based, following a latency-insensitive protocol \cite{Fleming13_lic_thesis}.
To analyze which values should be computed by which decoupled unit and which values should be communicated, we use the def-use chain encoded in the SSA form of the code (each SSA value usage can be traced to its unique definition \cite{ssa_book}).
We follow a standard approach to automatically generate a DAE architecture \cite{szafarczyk_fpl23}:
\begin{enumerate}
    \item \textbf{AGU:} Each memory operation to be decoupled is changed to a \texttt{send\_address} FIFO write that sends the memory address to the DU. 
    \item \textbf{CU:} Dually, in the CU each memory operation to be decoupled is changed to a \texttt{consume\_value} or \texttt{produce\_value} FIFO read function that receive or send values to or from the DU.
    \item \textbf{Dead code elimination (DCE):} We apply DCE in the CU to remove any unnecessary address generation code. In the AGU, we delete side effect instructions that are not part of the address generation def-use chains, and then also apply DCE followed by control-flow simplification to remove redundant basic blocks.
\end{enumerate}


\subsection{Static Loop Fusion} \label{sec:compiler_prelimenaries}


Polyhedral compilers represent memory operations inside loop nests as integer sets \cite{girbal2006semi, pldi_08_Bondhugula_poly, polly_ppl}:
\begin{enumerate}
    \item The \textit{domain} set describes the set of loop iterations in which a statement is executed. 
    \item The \textit{schedule} set maps domain elements to a point in time. Given two schedule instances, we can determine which one comes first in program order.
    \item The \textit{access} set maps domain elements to a point in space, representing the accessed memory location.
\end{enumerate}
For example, the domain ($D$), schedule ($S$), and access ($A$) functions of the \textit{i}-loop store $st_A$ and \textit{j}-loop load $ld_A$ in figure~\ref{fig:Pipeline_a} are:
\begin{align*}
& D_{st_A} = \{st_A[i]: 0 \leq i < N\},&& D_{ld_A} = \{ld_A[j]: 0 \leq i < M\} \\
& S_{st_A} = \{st_A[i] \rightarrow [0, i]\},&& S_{ld_A} = \{ld_A[j] \rightarrow [1, j]\} \\
& A_{st_A} = \{st_A[i] \rightarrow f(i)\},&& A_{ld_A} = \{ld_A[j] \rightarrow g(j)\}
\end{align*}
The set intersection of two access relations can be used to find dependencies between the two corresponding operations.

Static loop fusion for the code in figure~\ref{fig:Pipeline_a} can be expressed as a transformation on the schedule of the load: $\mathcal{T}_{Fusion} = \{[1, j] \rightarrow [0, i]\}$ (together with transformations to account for $N \neq M$).
$\mathcal{T}_{Fusion}$ might introduce a new dependency between the store and load.
The transformation is only legal if the dependency distance of the new dependency is non-negative: this implication has to hold $A_{st_A}[k] = A_{ld_A}[l] \implies k < l$, where $k, l$ are some iterations in the fused loop.
In other words, if in the original program a given store writes to an address that a given load later uses, then in the fused loop the store must execute in an earlier iteration than the load.

The legality of loop fusion can be reduced to checking the legality of pairwise loop permutation \cite{popl_fusion}---the permutation should not break dependencies.
However, if the address expressions do not form affine functions, then the legality check does not have enough information about dependency distances to be useful.
One can over-approximate non-affine functions as affine \cite{poly_cc}, but this does not help in all cases, e.g., over-approximation can introduce spurious dependencies on codes with data-dependent addresses.
Our dynamic loop fusion is more lenient, requiring only monotonically non-decreasing addresses. 
However, we stress that our aim is not to replace the polyhedral approach to static loop fusion. 
Clearly, static loop fusion is preferable whenever possible, especially since it can be combined with other transformations in one framework \cite{popl_fusion}.
Rather, we aim to enable fusion in cases where static approaches are fundamentally infeasible.

\section{Address Monotonicity} \label{sec:address_monotonicity}

We now describe the concept of address monotonicity in more detail and contrast it with affine addresses.

\subsection{Motivation for Monotonicity}

Assume that we have a memory dependency across loops.
If we can prove at compile time that the address of the dependency source is monotonically non-decreasing, then at runtime the loop with the dependency destination only has to check if the address it accesses is lower than the most recently accessed address in the source loop---the dependency destination does not need to see the full history of memory accesses made in the other loop.
This paves the way for our efficient hardware dynamic memory disambiguation across loops described in section \ref{sec:dataUnit}.
We now describe how addresses can be proven to be monotonically non-decreasing.

\subsection{Monotonic Chain of Recurrences}

Compilers can represent expressions inside loops as a \textit{Chain of Recurrences} (CR) \cite{scev_bachmann, scev_engelen, scev_pop_cohen}:
\begin{equation*}
\{base, \odot, step\},
\end{equation*}
where $base$ and $step$ can themselves be a CR, and $\odot \in \{+, \times, \div\}$.
To reason about memory addresses, we typically use the constraints: $base, step \in \mathbb{N}$ if they are not a CR; and $\odot = \{+, \times\}$.
Both LLVM and GCC provide a CR analysis called Scalar Evolution (SCEV) \cite{scev_gcc_2004, scev_pop2004fast}.

A CR is \textit{affine} iff it is an add recurrence and iff its step is a constant expression not containing any CRs \cite{polly_ppl}. 
A CR is \textit{monotonically non-decreasing} iff its step is non-negative \cite{monotonic_scev}.
For brevity, we use the term \textit{monotonic} to mean monotonically non-decreasing in the rest of the paper.

Monotonic CRs are more general than affine CRs and handle control flow better \cite{monotonic_scev}.
For example, the CR of a row-major $N \times N$ matrix traversal is affine and monotonic: $\{\{0, +, N\}, +, 1\}$.
But the CR for an FFT traversal is not affine anymore, only monotonic: 
$\{\{0, +, 1\}, +, \{2, \times, 2\}\}$.

An address expression is monotonic w.r.t. a given loop depth iff the loop CR expression consists of only monotonic CRs.
Monotonically non-increasing addresses (i.e., using $step \in \mathbb{Z}$ and adding $\div$ to $\odot$) can also be supported by just flipping signs in the hazard detection logic, but we do not discuss this further in this paper. 



\subsection{Monotonicity in Sparse Array Formats}
Data-dependent accesses cannot be analyzed using the CR formalism, yet their underlying access pattern is often monotonic.
For example, sparse matrix formats, like CSR, produce address sequences that retain the partial order of the original row-major matrix traversal.
Other data-dependent accesses that are not monotonic by definition can be made monotonic with pre-sorting.
To support dynamic loop fusion on these codes, we allow the user to annotate memory operations asserting that the address is monotonic in a given loop.

\subsection{Non-Monotonic Outer Loops} \label{sec:isMaxIterNeeded}

We require a monotonic CR for the innermost loop of the memory dependency source; the outer loop CRs can be non-monotonic.
Consider this producer-consumer example:

\begin{verbatim}
    for (i=0; i<ITERS; ++i)
        for (j=0; j<N; ++j) 
            store A[j];
    for (k=0; k<M; ++k) 
        load A[k];
\end{verbatim}
The store innermost $j$-loop is monotonic, but the outer $i$-loop is not---advancing the $i$-loop causes the store address to reset.
We encode this information in our schedule (section \ref{sec:schedule}), so that in this case our DU will know that it has to wait for the last $i$-loop iteration to be sure that a given $A[j]$ store address in the $j$-loop will not be repeated.


\subsubsection{Detecting Non-Monotonicity}

Given an address expression $f(i_1, i_2, ..., i_n)$ nested within $n$ loops (where $n$ is the innermost loop depth), a $k, 1 \leq k < n$ loop depth is \textit{non-monotonic} if there exists a $j > k$ loop depth such that $CR_k.step < (CR_j.step \times tripCount_j)$, where $CR_k.step$ is the step component for loop $k$, and $tripCount_j$ is the number of times loop $j$ executes.
In other words, a given outer loop $k$ is non-monotonic if there exists a deeper nested loop whose entire execution contributes a larger value to the address value than one $k$-loop iteration.
A $CR_k$ for loop $k$ might not exist, in which case that loop depth is trivially marked as non-monotonic.

For example, the outer loop in a row-major $N \times M, N > 1, M > 1$ matrix traversal is monotonic, because its step is $M$, which is not lower than $CR.step \times tripCount=M$ of the inner loop.
On the other hand, the outer loop in a column-major traversal is non-monotonic, because its step value is $1$, which is lower than $CR.step \times tripCount=M \times M$ of the inner loop.

The above expressions are usually symbolic.
We substitute symbols with their maximum values (after a value range analysis).
This makes our monotonicity checks conservative---we might get false positives, but never false negatives.
The checks could be performed at runtime instead, which would make the result precise.
However, false positives did not occur in our evaluation, so we leave this for future work.

\section{Program-Order Schedule for Hardware} \label{sec:schedule}

Our schedule representation allows multiple loops to run in parallel, as opposed to being sequentialized as in existing dynamic memory disambiguation approaches for HLS \cite{lsq_2014, Josipovic_Brisk_Ienne_2017, szafa_fpt23, unified_memory_dependency_spec_hls, dai_speculative_data_hazards}.
Section \ref{sec:compiler_prelimenaries} discussed the schedule representation used in polyhedral compilers.
We use a similar representation at runtime, but with the following optimizations for hardware:
\begin{enumerate}
    \item Each loop depth is represented by one element in the schedule tuple, instead of a multi-dimensional point.
    \item Each schedule element is incremented by 1 for each invocation of the loop body corresponding to that element---no dependencies between schedule elements are introduced across loops.
    Repeated invocations of inner loops do not cause the corresponding schedule elements to wrap around.
    \item Schedule comparisons between two operations involve just one comparison between the schedule elements corresponding to the innermost shared loop depth of the operations, as opposed to comparing whole tuples as is the case in the polyhedral schedules.
\end{enumerate}

Consider these two nested loops for example:
\begin{verbatim}
    for (i=0; i<N; ++i) 
        for (j=0; j<2; ++j) 
            ld_0; st;
        for (k=0; k<4; ++k) 
            ld_1;
\end{verbatim}
Our DAE pass will decouple this code into two loop PEs:
\begin{multicols}{2}
\setlength{\columnseprule}{1pt}
\def\columnseprulecolor{\color{gray}}
\begin{verbatim}
for (i=0; i<N; ++i) 
    for (j=0; j<2; ++j)
        ld_0; st;
\end{verbatim}
\columnbreak
\begin{verbatim}
for (i=0; i<N; ++i) 
    for (k=0; k<4; ++k) 
        ld_1;
\end{verbatim}
\end{multicols}
\noindent
Assume $i=1, j=0$ for the left PE; and $i=0, k=3$ for the right PE.
The $st$ schedule will be $\{2, 3\}$; the $ld_1$ schedule will be $\{1, 4\}$.
To check if a $st$ schedule instance comes before a $ld_1$ schedule instance in program order, written as $schedule_{st} \prec schedule_{ld_1}$, we compare the schedule elements corresponding to the $i$-loop.
Similarly, to check $schedule_{st} \prec schedule_{ld_0}$, we compare the $j$-loop schedule elements.

The below table shows the difference in evolution of our and the polyhedral schedule representation for the $st$ operation:

\vspace{1em}
\begin{tabular}{r | l  l  l  l}
\hline
iters: & i=0, j=0 & i=0, j=1 & i=1, j=0 & i=1, j=1 \\
poly: & \{0, 0, 0, 1\} & \{0, 0, 1, 1\} & \{1, 0, 0, 1\} & \{1, 0, 1, 1\} \\
ours: & \{1, 1\} & \{1, 2\} & \{2, 3\} & \{2, 4\} \\
\hline
\end{tabular}
\vspace{1em}

\noindent
The additional dimensions in the polyhedral schedule are used to represent program order within loops.
How can we avoid the additional dimensions in our schedule and still recover program order within loops?
For example, we want to know that $schedule_{ld_0} \prec schedule_{st}$ even when both schedules will be equal to $\{2, 3\}$.
Our insight is to \textit{configure the schedule comparator} based on the topological order of memory operations in the program.
In a $schedule_{ld_0}[1] \odot schedule_{st}[1]$ comparison, where the index 1 refers to the $i$-loop, we will configure $\odot=\, \leq$.
Dually, to check $schedule_{st} \prec schedule_{ld_0}$, we would synthesize: $schedule_{st}[1] < schedule_{ld_0}[1]$.

In summary, our compiler pass statically configures schedule comparators used in the DU for each dependency pair, so that we can recover total ordering without additional schedule dimensions and without the need to compare entire schedule tuples.



\subsection{Integration of Non-Monotonic Outer Loops} \label{sec:marking_last_iter}
For each non-monotonic outer loop $k$, we add a $lastIter$ bit to the schedule that will be set in the AGU if the corresponding request was generated on the last $k$-loop iteration.
Our DU uses $lastIter$ bits as hints to expedite disambiguation---they are not essential for correctness.
Non-monotonic loops for which $lastIter$ bits cannot be generated are still supported.

\subsection{Schedule Generation in AGUs} \label{sec:schedule_generation}

Our compiler adds schedule-generating instructions for each AGU memory request as follows:
\begin{enumerate}
    \item At the start of the AGU, an $n$-tuple $schedule$ is initialized to 0, where $n$ is the request loop depth.
    \item At each loop depth $1 \leq i \leq n$, a $schedule[i]$ increment instruction is inserted to the beginning of the first non-exiting basic block of the $i$-loop body.
    \item For each non-monotonic loop $k$, we add a $lastIter[k]$ comparison instruction that evaluates to true if this is the last $k$-loop iteration. This involves calculating loop predicates one iteration in advance. 
    The $lastIter$ bit is just a hint and is set to false if the loop predicate cannot be calculated one iteration in advance.
    \item At the end of the AGU, each $schedule$ element is set to a sentinel value that signals to the DU that there will be no more requests from this AGU. 
\end{enumerate}
Schedules are implemented in 32-bit registers and are shared between all memory operations in the same AGU.
Future work could use range analysis to decrease schedule bit sizes.



\section{Data Unit with Hazard Detection} \label{sec:dataUnit}
\begin{figure}[t]
\centering

\centering
\includegraphics[width=0.43\textwidth]{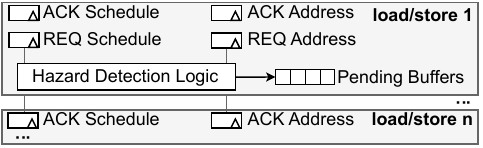}

\caption{Data Unit (DU) consisting of of $n$ Load Store Units.}

\label{fig:LSUs}
\end{figure}

Each program base pointer that has unpredictable dependencies, or that has dependencies across loops that cannot be fused statically, is assigned its own DU to perform dynamic disambiguation.
Figure~\ref{fig:LSUs} shows a high-level DU organization.
In our implementation, each program load and store gets its own port; future work could study port sharing.

Each load and store keeps track of the address and schedule corresponding to the most recent ACK received from, and the next request to be sent to, the memory controller.
It also has buffers to hold addresses, schedules, and values (in case of stores) for pending requests (not yet ACKed requests) .

The hazard detection logic compares the address and schedule of its next request with the address and schedule of the most recent ACK of its dependency sources.
The next request will only be sent to the memory controller and moved to the pending buffer if the check succeeds.
The check and enqueueing logic is spread across multiple pipeline stages---there is no negative load latency impact, because, thanks to the DAE architecture, load addresses run ahead of load consumers giving us ample cycle budget.
The pending buffers are implemented in registers to enable associative searching needed for store-to-load forwarding (section \ref{sec:store_to_load})---their size depends on the DRAM burst size.

In the rest of this section, we describe how the monotonicity property and our schedule representation are used to enable dynamic memory disambiguation across loops.

\subsection{Hazard Detection Problem Statement} \label{sec:problem_statement}

We are trying to check if a memory operation $a$ has a data hazard with memory operation $b$.
Assume $a$ is nested in $n$ loops, $b$ is nested in $m$ loops, and they both share a loop at depth $k, k \leq n, k \leq m$.
Informally, given a $req.schedule_a$ and $req.address_a$ corresponding to the next $a$ request, and $ack.schedule_b$ and $ack.address_b$ corresponding to the most recent ACK for operation $b$, our hazard detection logic deems the next $a$ request safe if either of the two conditions holds:
\begin{enumerate}
    \item The next $a$ request comes before the most recent $b$ ACK in program order.
    \item The next $a$ request comes after the most recent $b$ ACK in program order, but $req.address_a$ will no be accessed by operation $b$ in the the $(ack.sdchedule_b,$ $req.schedule_a)$ range.
\end{enumerate}

We now describe each of these points in more detail, before composing the equations implementing these two checks into a general \ref{eq:hazard}.
In the following discussion, we use the term  ``$(schedule_a, schedule_b)$ time range'' to mean the sequence of memory requests $b'$ such that $schedule_a[k] < schedule_{b'}[k] < schedule_{b}[k]$, where $k$ is the innermost common loop depth of operation $a$ and $b$.
We use open parenthesis and box brackets to represent open and closed intervals, respectively.

\subsection{Comparing Schedules} \label{sec:compare_schedules}

If operations $a$ and $b$ do not share any loops ($k=0$), then the relative schedule program order will always match their topological program order and we do not need to synthesize any comparisons.
Otherwise, if the shared loop depth $k > 0$, we synthesize the following comparison to check if the next $a$ request comes before the most recent $b$ ACK:
\begin{align*}
\tag{Program Order Safety Check} \label{eq:schedule} \\
& req.schedule_{a}[k] \odot ack.schedule_{b}[k] \parallel \\
\bigl( &req.schedule_{a}[k] \odot req.schedule_{b}[k]\,\, \&\,\, noPendingAck_b \bigr)
\end{align*}
Where $\odot = \leq$ if $a \prec b$ in topological program order, else $\odot = <$.
The $noPendingAck$ term is a single bit that is set if $b$ is not waiting for any ACKs.
The second equation line makes sure that the $a$ request is deemed safe if there are no further $b$ requests in the $[ack.schedule_b,$ $req.schedule_{a})$ time range.

Since we only use the schedule element corresponding to the innermost shared loop of the two memory operations, we do not need to synthesize the rest of the schedule.

\subsection{Checking Address Reset in Schedule Range} \label{sec:address_monotonicity_in_range}

If the above check fails, then for request $a$ to be safe we check that operation $b$ will not access $req.address_a$ in the $(ack.schedule_b,$ $req.schedule_{a})$ time range.
If all operation $b$ loop depths are monotonic, this is a simple $req.address_a < ack.address_b$ check.
If some $b$ loops are non-monotonic, we need to guarantee that $ack.address_b$ will not be reset in the considered schedule range:
\begin{alignat*}{2}
\tag{No Address Reset Check} \label{eq:addr_reset} \\
lastIterCheck\,\, \&\,\, req.schedule_a[l] = ack.schedule_b[l] + \delta
\end{alignat*}
Here, $\delta = 1$ if $a \prec b$, else $\delta = 0$; $l$ is the deepest non-monotonic loop depth in the $b$ operation loop nest such that $l \leq k$; and the $lastIterCheck$ term is an \texttt{AND}-reduction of the $b$ $lastIter$ bits:
\[
ack.lastIter_{b} = (bit_1, ..., bit_k, \underbrace{bit_{k+1}, ..., bit_{m-1}}_\text{AND-reduction}, bit_{m}),
\]
where $bit_j, 1 \leq j \leq m$ is set to 1 at compile time if the $j$ loop is monotonic and thus optimized away from the reduction; otherwise $bit_j$ will be set dynamically on the last iteration of the $j$ loop according to the procedure from section \ref{sec:marking_last_iter}.

The first term in the \ref{eq:addr_reset} guarantees that all non-monotonic child loops of $k$ are on their last iteration, and thus will not reset the $b$ address.
The second term guarantees that the $b$ address will not reset as a result of advancing in some parent loop of $k$.
Only bits corresponding to non-monotonic loop depths are considered in the \texttt{AND}-reduction.
Similarly, if all $[1, k]$ loops are monotonic, then the second term is omitted.

\subsubsection{Example}
Consider the following code:
\begin{verbatim}
  for (; a < A; ++a)       // depth 1: non-monotonic
    for (; b < B; ++b)     // depth 2: monotonic
      for (; c < C; ++c)   // depth 3: non-monotonic
        for (; e < E; ++e) // depth 4: monotonic
          mem_op_b;
      for (; d < D; ++d)   // depth 3
        mem_op_a;
\end{verbatim}
Here, the $b$ address is non-monotonic at loop depth $1$ and $3$.
The innermost common loop depth of operations $a$ and $b$ is $k=2$.
The innermost non-monotonic $b$ loop depth that is lower than $k$ is $l=1$.
Thus, the \ref{eq:addr_reset} checks $req.schedule_{a}[1] = ack.schedule_{b}[1]$ to guarantee that $b$ will not have any more $l$-loop iterations until reaching the $req.schedule_{a}$ point.
And it will check if $ack.lastIter_{b}[3]$ is set to guarantee that the $b$ address will not reset by advancing in the non-monotonic $3>k$ loop.

\subsection{Hazard Safety Check} \label{sec:hazard_safety_check}

With the ability to compare program order schedules and guaranteeing that addresses do not reset in a given schedule range, we can now construct a general data hazard check.
The next $a$ request is safe to execute w.r.t the most recent $b$ ACK if:
\begin{align*}
\tag{Hazard Safety Check} \label{eq:hazard} \\
& \ref{eq:schedule} \parallel \\
& \bigl( req.address_a < ack.address_b\,\, \&\,\, \ref{eq:addr_reset} \bigr)
\end{align*}

\subsubsection{Complexity} \label{sec:hazard_complexity}
The \ref{eq:hazard} simplifies to just one $req.address_a < ack.address_b$ comparison if $a$ and $b$ do not share loops. 
If $b$ has non-monotonic loops, then the \ref{eq:addr_reset} adds at most one \texttt{AND} reduction and one equality check.
The number of comparisons grows to three if there is a shared loop thanks to the \ref{eq:schedule}.
In general, given a program with $n$ operations, if we check every possible dependency pair, then the number of comparisons is $\mathcal{O}(n^2)$---reducing complexity becomes important as the number of loads and stores grows.
Loads do not have to check for hazards against other loads.
Also, WAR checks where the written value depends on the read value can be omitted, as previous work has already pointed out \cite{josipovic_shrink_or_shed}.

\begin{figure}[t]
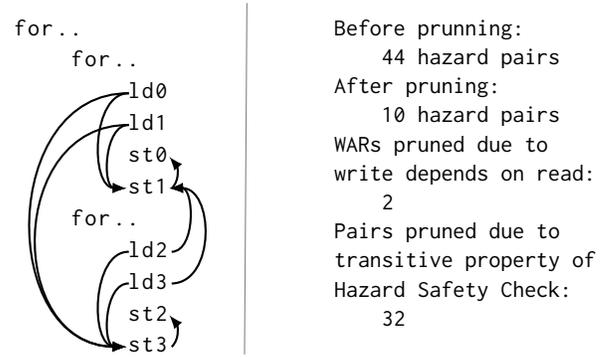

\centering

\begin{multicols}{2}
\begin{lstlisting}[basicstyle=\ttfamily, escapeinside={(*}{*)}]
  for..
      for..
          (*\tikzmark{ld0_d}*)ld0(*\tikzmark{ld0}*)
          (*\tikzmark{ld1_d}*)ld1(*\tikzmark{ld1}*)
          (*\tikzmark{st0_d}*)st0(*\tikzmark{st0}*)
          (*\tikzmark{st1_d}*)st1(*\tikzmark{st1}*)
      for..
          (*\tikzmark{ld2_d}*)ld2(*\tikzmark{ld2}*)
          (*\tikzmark{ld3_d}*)ld3(*\tikzmark{ld3}*)
          (*\tikzmark{st2_d}*)st2(*\tikzmark{st2}*)
          (*\tikzmark{st3_d}*)st3(*\tikzmark{st3}*)
\end{lstlisting}    
\DrawArrow[black, thick, in=-50]{st1}{st0}
\DrawArrow[black, thick, in=-50]{st3}{st2}
\DrawArrow[black, thick, bend right=90]{ld2}{st1}
\DrawArrow[black, thick, bend right=90, looseness=1.2]{ld3}{st1}
\DrawArrow[black, thick, bend right=90]{ld0_d}{st1_d}
\DrawArrow[black, thick, bend right=90]{ld1_d}{st1_d}
\DrawArrow[black, thick, bend right=90, looseness=1.3]{ld0_d}{st3_d}
\DrawArrow[black, thick, bend right=90, looseness=1.4]{ld1_d}{st3_d}
\DrawArrow[black, thick, bend right=90]{ld2_d}{st3_d}
\DrawArrow[black, thick, bend right=90]{ld3_d}{st3_d}
\DrawArrow[gray,-]{[xshift=1cm, yshift=1.2cm]ld0}{[xshift=1cm,yshift=-0.1cm]st3}
\columnbreak
\begin{verbatim}
Before prunning:
    44 hazard pairs
After pruning:
    10 hazard pairs
WARs pruned due to 
write depends on read:
    2
Pairs pruned due to
transitive property of 
Hazard Safety Check:
    32          
\end{verbatim}
\end{multicols}
\vspace{-1cm}
\caption{Result of pruning hazard pairs in the later evaluated FFT code. Each memory operation checks for safety against at most one operation per loop depth (e.g., $ld_0$ checks against $st_3$ in its first loop depth, and against $st_1$ in the second).}

\label{fig:Prunning}
\end{figure}

However, by exploiting the transitive property of our \ref{eq:hazard} we can prune many more hazard pairs.
Assume that we have three memory operations with the following topological program order $c \prec b \prec a$. 
The safety check of $a$ against $c$ can be omitted, since $a$ already checks against $b$, and $b$ checks against $c$.
Operation $c$ still has to be checked against $a$ if there is a CFG path via a loop backedge from $a$ to $c$.
With pruning, the worst case number of comparisons reduces to $\mathcal{O}(nd)$, where $d$ is the maximum loop depth.
For example, in the an FFT code which we later evaluate, the above pruning procedure decreased the number of hazard safety checks from 44 to 10 (32 checks were pruned due to our transitivity property, 2 due to a store to load dependency).
Figure~\ref{fig:Prunning} shows the result of such pruning.  

\subsection{Store-to-Load Forwarding} \label{sec:store_to_load}

We support store-to-load forwarding by allowing loads to directly access values from a dependent store's pending buffer.
We specialize the \ref{eq:hazard} for RAW dependencies: instead of using the address and schedule of to the most recent store ACK, we use the address and schedule of the next store request.
In addition, we perform an associative search of the pending store buffer, using the load address as a key.
If the modified RAW check succeeds, then the dependent value will either already have been committed and ACKed, or it is in the store pending buffer and our associative search will find it.
Hits from the buffer search can be used by the load directly, without issuing a DRAM request.
If there are multiple values with the same address in the pending buffer, the youngest is chosen (this is cheap to implement in FIFO buffers).


The case where two stores that can both forward a value with the same address to the same load is impossible.
Assume the following program order of operations that all use the same address: $store_0 \prec store_1 \prec load$.
The $store_1$ will not be able to move its value to its pending buffer until after the $store_0$ value has been ACKed---its WAW hazard detection will stall it.
Conversely, the $load$ will not use the $store_0$ value, because it will stall on the RAW check against $store_1$---the $load$ will wait for $store_1$ to move its value to its pending buffer.

With forwarding, some WAW checks cannot be pruned anymore, because load RAW checks do not use store ACKs.
In our above example, if all operations are in the same loop, then the $store_0$ WAW check against $store_1$ cannot be pruned, because the $load$ ACK might be updated as a result of store forwarding from $store_1$, with the forwarded value not yet ACKed in $store_1$. 


\subsection{Intra-Loop RAW Hazards}

A timely disambiguation of RAW hazards, where both the load and store are in the same loop PE, is crucial since any unnecessary stalls would be repeated on every iteration, resulting in a large throughput reduction.
As our evaluation in section \ref{sec:Evaluation} will show, store-to-load forwarding becomes crucial in intra-loop RAW dependencies.

In addition to forwarding, there is another term needed in the RAW \ref{eq:hazard} to make intra-loop RAW hazard checks timely. 
Consider this simple code:
\begin{verbatim}
    for (i = 0; i < N; ++i)
        d = data[i];
        data[i] = work(d);
\end{verbatim}
The load and store address distribution is $\{0, 1, 2, ...\}$---there is no actual RAW hazard, but assume that we do not know this at compile time.
In this situation, the RAW \ref{eq:hazard} for a given load at iteration $k$ will only succeed once the next store request in the DU is for iteration $k-1$ and there are no outstanding store ACKs. 
If the next store request is for an earlier iteration, e.g., an earlier store request is waiting for its store value, then the load would have to be stalled, even though it would be perfectly safe to execute it.

We solve this issue by adding a $NoDependence$ single-bit term to the RAW \ref{eq:hazard}.
For each intra-loop RAW hazard pair, $NoDependence$ is set in the AGU to the result of $req.address_{load} > req.address_{store}$, where $req.address_{load}$ is the next load address to be sent to the DU, and $req.address_{store}$ is the most recent store address that was sent to the DU.
When $NoDependence$ is true, and the \ref{eq:addr_reset} evaluates to true, then the load can be deemed safe since the monotonicity property implies that all store addresses up to $req.schedule_{load}$ are lower than $req.address_{load}$.

Note that a similar check is not needed for intra-loop WAW dependencies, since stores do not stall the datapath if sufficient buffering is provided for the store values.

\begin{figure}[t]
\centering

 \begin{subfigure}[b]{0.11\textwidth}
     \centering
\begin{verbatim}
for ...
  if (cond)
    store;
  load;

  
\end{verbatim}%
\caption{Store under \textit{if}-condition.}
     \label{fig:Speculation_a}
 \end{subfigure}%
 \hfill%
 \begin{subfigure}[b]{0.15\textwidth}
     \centering
     \includegraphics[width=\textwidth]{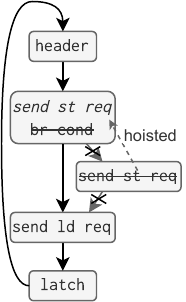}
     \caption{AGU speculates requests.}
     \label{fig:Speculation_b}
 \end{subfigure}%
 \hfill%
 \begin{subfigure}[b]{0.15\textwidth}
     \centering
     \includegraphics[width=\textwidth]{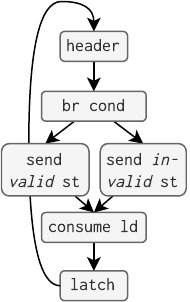}
     \caption{CU kills mis-speculations.} 
     \label{fig:Speculation_c}
 \end{subfigure}
 
\caption{Memory requests in \textit{if}-conditions are speculated.}
\label{fig:Speculation}
\end{figure}

\section{Handling Control Flow}

The \ref{eq:hazard} relies on the ability of the DU to detect that a given memory operation has completed a certain schedule time range or a certain address range.
This assumes that AGUs supply an operation's schedule and address for every loop iteration.
This assumption is broken by operations inside \textit{if}-conditions, which can lead to a deadlock.
Consider the code in figure~\ref{fig:Speculation}(a).
If the \textit{if}-condition in this loop is never true, then the store will never update its ACK address and schedule, and thus the RAW \ref{eq:hazard} in the DU would never succeed.
Eventually, the AGU would fill the load request FIFO, resulting in a deadlock.

This could be avoided by using separate AGUs for each memory operation---the store AGU would be guaranteed to at least send a final sentinel value, which would eventually cause the RAW hazard check to succeed.
However, this would again mean that some loops need to run to completion before the check can be performed.

A better approach is to \textit{speculatively} send memory requests.
We adapt the work presented in \cite{szafa_cc25} to implement speculation in a DAE architecture.
In our example, the store request can be hoisted out of the \textit{if}-condition in the AGU.
Then, the store values going to the DU from the CU can be tagged with a \textit{valid} bit that signals if the value should be committed or not, depending on the actual control flow at runtime.
Figure \ref{fig:Speculation} shows the AGU and CU control-flow graphs that implement such speculation.

Previous work used speculation to remove loss-of-decoupling (LoD) problems in DAE architectures \cite{szafa_fpt23, desc_cpu, loss_of_decoupling}.
A LoD arises when the AGU has dependencies on values that have to be loaded from a DU or calculated by a CU, preventing the AGU from running ahead \cite{effectiveness_of_decoupling_sc}.
Our approach is the same as previous work, but we apply it to all \textit{if}-conditions with the goal of producing an $(address, schedule)$ pair for each loop iteration in the AGU.
As a side benefit, speculation also makes us immune to the control-dependency LoD problem.

Mis-speculated loads are executed normally in the DU.
The read in the CU CFG is moved to the same location where it was speculated in the AGU.
This guarantees that the order of load requests made from the AGU is the same as the order of load value consumption in the CU, on every CFG path.
After reading a speculated load value, the CU can simply not use it if it takes a CFG path where the load value is not needed.
Since the basic block location of the speculated load value consumption changes, we also need to adjust any $\phi$-nodes that use the load value.

\begin{figure}[t]
\centering

\includegraphics[width=0.46\textwidth]{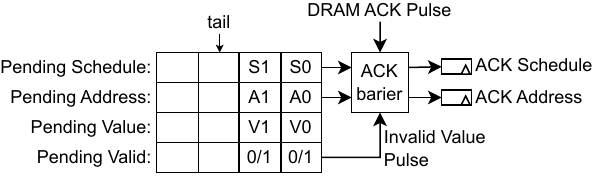}
\caption{Handling of mis-speculated stores in the DU. Before being moved to the pending buffer, invalid stores are also checked for safety to uphold the transitive property of the \ref{eq:hazard}. They do not submit DRAM requests. When reaching the head of the pending buffer, they update the ACK registers without having to wait for an ACK.}

\label{fig:PendingBuffers}
\end{figure}

Mis-speculated stores are detected using the \textit{valid} bit in store values coming from the CU.
Invalid stores are never committed to memory---there is no need for costly rollbacks.
However, invalid stores should eventually update the ACK registers to signal that a given time and address range was completed by the store.
Figure~\ref{fig:PendingBuffers} shows our approach to this.

If a whole loop with memory operations is under an \textit{if}-condition, then we fold the \textit{if}-condition into the loop body and execute the whole loop speculatively.
This was not a performance problem in our evaluation, but future work could investigate a whole loop speculation scheme that does not require executing all loop iterations.

\section{Evaluation} \label{sec:Evaluation}

\begin{figure}[t]
\centering

\includegraphics[width=0.47\textwidth]{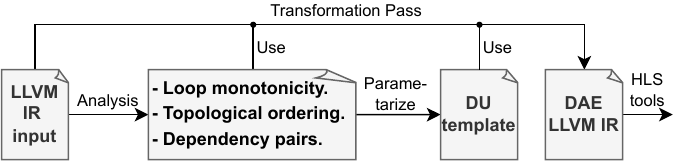}
\caption{Our compiler/hardware co-design flow. We use the Intel HLS tool in this paper. Our DU is parametrized by the number of loads and stores. The DU disambiguation logic is parameterized for each hazard pair (dependency source and destination) based on the loop nest monotonicity of the dependency source; and the relative topological ordering of the dependency source relative to the destination.}

\label{fig:ToolFlow}
\end{figure}

We implemented our compiler/hardware co-design in the Intel HLS compiler \cite{sycl_fpga_compiler}.
Figure~\ref{fig:ToolFlow} shows our tool flow.
Our implementation and evaluation are publicly available \cite{zenodo_dynamic_fusion}.

\begin{table*}[!t]
\centering

\renewcommand{\arraystretch}{1}

\caption{Performance, area usage, and circuit frequency of the \texttt{STA}, \texttt{LSQ} \cite{szafarczyk_fpl23}, \texttt{FUS1}, and  \texttt{FUS2} approaches. The second column reports the number of PEs and DUs generated by our \texttt{FUS} approach, together with loads and stores per DU.}

\label{tab:benchmarkTable}
\centering
\begin{tabular}{l | cccc | cccc | cccc | cccc }
\hline
\multirow{2}{*}{\textbf{Kernel}} & \multicolumn{4}{c|}{\textbf{Number of}} & \multicolumn{4}{c|}{\textbf{Area in 1000s of ALMs}} & \multicolumn{4}{c|}{\textbf{Freq in MHz}} & \multicolumn{4}{c}{\textbf{Time in seconds}} \\

& PE & DU & LD & ST
& STA & LSQ & FUS1 & FUS2
& STA & LSQ & FUS1 & FUS2
& STA & LSQ & FUS1 & FUS2 \\
\hline
\hline

RAWloop & 2 & 1 & 1 & 1 & 78 & 79.6 & 82.5 & 83.3 & 304 & 268 & 263 & 239 & 6.8 & 33.3 & 3.9 & 4.4 \\
WARloop & 2 & 1 & 1 & 1 & 78.1 & 79.6 & 82.2 & 82.2 & 279 & 264 & 261 & 261 & 7.1 & 33.5 & 4.1 & 4.1 \\
WAWloop & 2 & 1 & 1 & 1 & 78.3 & 80.8 & 88.4 & 88.4 & 294 & 269 & 251 & 251 & 6.8 & 7.5 & 4.1 & 4.1 \\
bnn & 2 & 1 & 2 & 2 & 78.9 & 85.1 & 93.5 & 95.2 & 279 & 244 & 266 & 257 & 39.2 & 3.2 & 1.6 & 1.6 \\
pagerank & 3 & 2 & 2/1 & 2/1 & 81.5 & 87.8 & 114.1 & 115.2 & 262 & 237 & 246 & 246 & 35.7 & 0.8 & 1.6 & 0.7 \\
fft & 2 & 2 & 4/4 & 4/4 & 102.7 & 102.7 & 150.4 & 152.2 & 246 & 246 & 221 & 219 & 7.8 & 7.8 & 2.8 & 1.7 \\
matpower & 2 & 1 & 4 & 2 & 82.1 & 97.6 & 105.4 & 108.6 & 274 & 193 & 260 & 257 & 18 & 3.7 & 12.3 & 1.6 \\
hist+add & 3 & 2 & 2/2 & 1/1 & 79.2 & 87.9 & 97.0 & 99.3 & 286 & 220 & 282 & 270 & 3.9 & 1 & 0.2 & 0.2 \\
tanh+spmv & 2 & 2 & 2/1 & 1/1 & 80.2 & 93.1 & 99.5 & 101.8 & 274 & 225 & 260 & 264 & 4.4 & 0.9 & 0.5 & 0.5 \\
\hline
\hline
\multicolumn{5}{r|}{Harmonic Mean:} & 1 & 1.07 & 1.22 & 1.24 & 1 & 0.86 & 0.92 & 0.9 & 1 & 0.12 & 0.1 & 0.07\\

\end{tabular}

\end{table*}

\subsection{Methodology}

We evaluate dynamic loop fusion on ten benchmarks where there is a possibility for parallelism across loops that is not exploited by current static and dynamic HLS tools.
All baselines use the Intel HLS compiler:
\begin{itemize}
    \item \texttt{STA}: baseline Intel HLS compiler performing automatic static loop fusion. This approach uses the same dynamically coalescing LSU as our DU. 
    \item \texttt{LSQ}: an implementation of dynamic scheduling within the Intel HLS compiler \cite{szafarczyk_fpl23}. An LSQ is used for memory accesses, but without support for dynamic coalescing. 
    This approach is representative of all current LSQ implementations in HLS \cite{lsq_2014, Josipovic_Brisk_Ienne_2017, szafa_fpt23, unified_memory_dependency_spec_hls, adaptive_loop_pipelining, dai_speculative_data_hazards}.
    \item \texttt{FUS1}: the dynamic loop fusion approach described in this paper, but with no store-to-load forwarding.
    \item \texttt{FUS2}: \texttt{FUS1} with store-to-load forwarding enabled.
\end{itemize}

We execute our benchmarks in hardware on the Altera Arria 10 GX1150 FPGA board \cite{pac_a10} with 2 banks of DDR4 memory (the memory controller uses two 512-bit channels). 
We use large datasets to ensure data is distributed across DRAM pages, resulting in variable latency.
Each code is executed three times and the minimum time is reported.
Area, reported as Adaptive Logic Modules (ALMs) \cite{alm}, and frequency are taken from Quartus 19.2 reports after place and route.
Our approach does not increase DSPs and BRAMs.

\subsection{Benchmarks}

We use irregular codes from dynamic HLS research \cite{benchmarks_jianyi_cheng_2019, josipovic_dynamatic_2022, szafarczyk_fpl23}, choosing codes where there are sibling loops that can benefit from our dynamic loop fusion.
For some benchmarks, we unroll outer loops to expose two inner loops that can be dynamically fused; or we compose multiple kernels to simulate applications composed of multiple tasks.
Some codes have address expressions that can be analyzed for monotonicity; some codes use data-dependent accesses that are asserted to be monotonic by the programmer.
We now list our benchmarks and the parameters used in this evaluation:
\begin{itemize}[]
    \item \textbf{RAWloop, WARloop, WAWloop:} each benchmark has two loops, each with one memory access, forming a RAW, WAR, or WAW dependency across loops. We use these benchmarks to compare our speedup to the maximum theoretical speedup. Complexity $\mathcal{O}(n)$. We set $n=10\,000\,000$.
    \item \textbf{bnn:} one layer of a sparse binarized neural network. There are two loops, both with data-dependent accesses that prevent fusion. We mark the inner loops as monotonic since we know that the sparse representation is monotonic. Complexity $\mathcal{O}(n^2)$. We set $n=10\,000$.
    \item \textbf{pagerank:} uses a compressed sparse row (CSR) format to iterate over the graph. Another two loops in the algorithm have a regular access pattern, but they cannot be fused because the irregular loop is between them. Complexity $\mathcal{O}(iters\,(nodes + edges))$. We set $iters=10$, $nodes=325\,729$, $edges=1\,497\,134$ using the \textit{web-NotreDame} graph \cite{snapnets}.
    \item \textbf{fft:} an FFT with the middle loop unrolled by a factor of two. 
    The non-affine accesses prevent loop fusion. 
    The \texttt{LSQ} and \texttt{STA} approach is equivalent for fft, because there are no hazards within loops that would need an LSQ.
    Complexity $\mathcal{O}(n\log_{}n)$. We set $n=1\,048\,576$.
    \item \textbf{matpower:} sparse matrix power using the CSR format with the outer loop unrolled by a factor of 2. Complexity $\mathcal{O}(nz^3)$, where $nz$ is the number of non-zero matrix values. We set $nz=4\,096$.
    \item \textbf{hist+add:} addition of two histograms. The \texttt{STA} approach can fuse the two histogram loops, but not the addition. Three $\mathcal{O}(n)$ loops. We set $n=10\,000\,000$.
    \item \textbf{tanh+spmv:} $tanh$ applied to a vector before it is used in a COO sparse matrix-vector multiplication. The $tanh$ loop has a store in an \textit{if}-condition, which we speculate. One $\mathcal{O}(n)$ loop followed by a $\mathcal{O}(nz)$ loop, where $nz$ is the number of non-zero matrix values. We set $n=10\,000$, $nz\simeq10\,000$.
    
\end{itemize}

\subsection{Results}

Table~\ref{tab:benchmarkTable} shows the area and performance results for the four approaches that we evaluate.
Dynamic loop fusion with forwarding is on average \textbf{14$\times$ faster than static HLS} and \textbf{4$\times$ faster than dynamic HLS} that uses an LSQ.

\subsubsection{Theoretical Speedup}
The RAW/WAR/WAW loop benchmarks have a theoretical speedup of $2\times$, but \texttt{FUS2} achieves a speedup of around $1.7\times$.
The lower speedup is due to the lower \texttt{FUS2} circuit frequency on these benchmarks.
The \texttt{LSQ} approach sees a slowdown relative to \texttt{STA} in the RAW/WAR loop benchmarks, because it cannot use a dynamically bursting LSU which stalls the load loop significantly (the LSQ used in \cite{szafarczyk_fpl23} uses a non-bursting LSU to gurantee that hazards are not violated \cite{szafa_fpt23}).
Store loops, e.g., WAWloop, do not suffer as much from a lack of bursting in the \texttt{LSQ} approach, because stores do not stall the LSQ pipeline. 

\subsubsection{Store-to-Load Forwarding Impact}
We observe that forwarding has no observable benefit on codes where the forwarding happens across loops, e.g., RAWloop.
This is expected in our evaluation setup, since without forwarding, the only penalty is an initial wait for the store ACK to be updated.
Forwarding across loops may become beneficial if the DRAM bandwidth becomes a bottleneck, which is likely to occur in practice once data parallelism is exploited.
Forwarding becomes crucial if the store and load are in the same loop and the dependency distance is lower than the store latency (e.g., fft, matpower, or pagerank).
Future work could use a more precise cost model and enable forwarding only where beneficial, e.g., always use forwarding for RAW dependencies inside loops, but for RAW dependencies across loops enable it only once the memory bandwidth is saturated.

\subsubsection{Which Codes Benefit from Dynamic Loop Fusion?\nopunct}

It only makes sense to fuse loops with similar time complexities.
Consider the pagerank benchmark as an example where fusion offers only a modest $1.1\times$ speedup over the \texttt{LSQ} approach.
The code consists of two $O(n)$ loops which go over graph nodes and one $O(n^2)$ loop which goes over edges.
Even if all three loops are fused, the runtime will still be dominated by the $O(n^2)$ loop.
We used the \texttt{web-Google} graph \cite{snapnets} with 875,713 nodes and 5,105,039 edges, which only has a theoretical speedup of $\approx1.3$ 
over \texttt{LSQ}.

We see the biggest benefit of using dynamic loop fusion in the ability to unroll outer loops of irregular codes without having to worry about breaking data dependencies (e.g., fft and matpower), and in the ability to perform task fusion at a fine-grained level (e.g., hist+add and tanh+spmv). 

\subsubsection{Area Overhead}
Dynamic loop fusion with forwarding comes at an average area increase of 24\% and frequency degradation of 9\% over static HLS.
The most area-hungry component is the dynamically coalescing LSU.
The \texttt{STA} approach also uses the costly coalescing LSUs, which amortizes the area overhead of fusion. 
The \texttt{LSQ} approach uses a simpler LSU, which explains its low area overhead.

For example, in the RAWloop benchmark, the \texttt{FUS2} DU consumes 1,550 ALMs (1,200 of which are dedicated to the pending buffers and its associative searching), whereas a single load LSU consumes 2,840 ALMs and the DRAM interconnect consumes 68,089 ALMs. 
If the OpenCL kernel runtime and DRAM interconnect are not counted, then our area overhead of dynamic loop fusion with forwarding increases to 2.1$\times$.
However, codes not using DRAM will not need the area budget for pending buffers, resulting in an overhead closer to what we report in table~\ref{tab:benchmarkTable}.


Hazard pairs pruning has a large impact on the area and critical path of codes with many loads and stores. 
For example, the FFT code uses two DUs, each with 4 loads and stores.
The unpruned FFT \texttt{FUS2} version uses 32\% more area and achieves a 28\% lower frequency than the pruned version.



\section{Limitations and Future Work} \label{sec:limit_and_future}

Our dynamic loop fusion approach can be integrated with common loop transformations used in HLS.
For example, loop tiling does not break the monotonicity of inner loops.
Dataflow designs---concurrent loops communicating via FIFOs--are automatically generated by our compiler given the program loop forest, as shown in Figure \ref{fig:LoopDecoupling}.
Loop unrolling---replicating the datapath of the inner loop---is also compatible with our DU, since we do not impose any limits on the number of memory ports.
However, our current implementation does not work with automatic loop unroll pragmas, and manual unrolling is needed instead.
This is because unrolling pragmas are typically implemented in the closed-source back-end of vendor compilers, and our compiler passes operate on LLVM IR in the middle-end.
We do not study unrolling in this work, because the types of irregular codes that we consider in this paper do not lend themselves to the same automatic parallelization as regular codes, e.g., due to unpredictable loop-carried dependencies.

In this work, we consider DRAM streaming applications, relying on a dynamically bursting and coalescing LSU to discover memory parallelism at runtime and on store-to-load forwarding to increase temporal locality.
In our current design, the amount of on-chip data reuse is limited by the size of the pending buffers, which have to be kept small to make associative searching feasible.
A cache memory hierarchy implemented in BRAM could further decrease the number of DRAM requests and increase temporal locality.
Recent work has advanced the state-of-the-art of non-blocking caches on FPGAs by storing Miss Status Holding Registers (MSHRs) in BRAM and using hash-based, instead of associative, searching \cite{mi_cache, crying_over_cache_hit_rate}.
In a DU with cache, the pending buffers could be changed to MSHRs with added schedule information and our store-to-load forwarding could be removed altogether, since temporal locality would be provided by the cache.

Using a BRAM-based cache and loop unrolling are orthogonal goals---supporting multiple memory ports is cheaper to do in BRAM than in DRAM, both in terms of the circuit area and available bandwidth.
However, since the automatic partitioning of BRAM into multiple banks cannot be performed for irregular code, a memory arbiter would have to be integrated to support multiple BRAM ports, as for example in \cite{cheng_bram_arbiter, lmc_leap_scratchpad, leap_shared_mem}.

\section{Related Work} \label{sec:related_work}

Our loop monotonicity analysis benefits from decades of research on abstract interpretation of recurrences \cite{ammarguellat1990automatic, monotonic_analysis_eigenmann, monotonic_analysis_padua, scev_bachmann, scev_engelen, scev_pop_cohen}.
Loop monotonicity has first been exploited in a practical setting by Gupta \textit{et al.} to synthesize race detection runtime checks in fork-join parallel programs \cite{loop_monotonic}.
However, they did not consider shared loops and non-monotonic outer loops.

We discussed the basics of the polyhedral compiler transformation framework in section \ref{sec:compiler_prelimenaries}, stating that codes with non-affine memory addresses or loop bounds cannot be analyzed.
Recent work has combined the Inspector/Executor (I/E) approach \cite{inspector_executor_91, inspector_executor_ferrante, inspector_executor_kennedy} with polyhedral transformations \cite{strout_spf, louis_polly_sc12, strout_cgo14_nonaffine_poly}.
The idea behind the I/E approach is to generate inspector code which gathers values of variables unknown at compile-time; and/or rearranges data structures in memory for better locality and to increase dependency distances.
The small overhead of the inspector code is offset by the throughput improvement obtained in the executor code.
For example, Strout \textit{el al.} proposed the Sparse Polyhedral Framework which uses ``uninterpretable functions'' to represent non-affine terms such as data-dependent memory accesses \cite{strout_spf}.
By proving basic properties about an uninterpretable function in the inspector code (e.g., monotonicity) a large amount of potential data dependencies can be ruled out, allowing the executor code to exploit more parallelism \cite{strout_sc16_monotonicity}.
Most recently, \cite{strout_sc23_sparse_fusion} proposed sparse fusion, an I/E technique that inspects the access patterns of multiple irregular loops and then creates an execution schedule that allows sibling loops to execute in parallel.
Although we share the same goal as these works, our approach is fundamentally different.
Instead of relying on inspector code to discover data dependencies, we resolve data dependencies ``on the fly'' in our DU.
We also provide a finer-grained monotonicity compiler analysis, discovering which specific loop depths cause an address expression to visit an earlier value, rather than deciding on the monotonicity of the entire loop nest expression. 

All previous work on dynamic memory disambiguation in HLS sequentializes loops that share a data dependency \cite{Josipovic_Brisk_Ienne_2017, straight_to_the_queue, szafa_fpt23, dac_runtime_dep_check_with_shiftreg, unified_memory_dependency_spec_hls}.
Cheng \textit{el al.} investigated compile time checks to prove that two loops do not access the same memory locations \cite{dynamic_inter_block_cheng}---their approach is the same as existing polyhedral optimizers, but uses a different formulation. 
Others have exploited the SCEV framework to augment the static analysis with dynamic checks in HLS \cite{offline_synthesis_of_online_dependence, liu_pipelining_polly, static_plus_dynamic_deps_trets}---these approaches are similar to multi-versioned SIMD CPU code, where the fast (SIMD) path is taken if a set of conditions evaluates to true at runtime.
All these works either only improve the throughput of single loops, or execute separate loops in parallel only if all iterations are independent.

Winterstein \textit{et al.} \cite{separation_logic_unrolling} used symbolic execution, based on separation logic, to prove the absence of aliasing when unrolling irregular loops into multiple PEs.
Later, they expanded the work to support aliasing limited to commutative operations by using locks to access a shared memory space \cite{separation_logic_unrolling_2}.
They rely on the commutative property because they cannot guarantee sequential consistency of accesses to the same memory spaces, as we have proposed here.

\section{Conclusions}

We have presented dynamic loop fusion, a compiler/hardware co-design approach that enables dynamic memory disambiguation across monotonic loops without the need for address history searches.
Our hazard detection logic is enabled by a novel program-order schedule representation, and by assuming monotonically non-decreasing addresses are in inner loops.
We have presented a compiler analysis, based on the chain of recurrences formalism, to detect loop monotonicity.
We have also shown that most codes contain addresses that are monotonic, making our approach applicable to a large class of applications.
On an evaluation of 10 irregular codes, dynamic loop fusion provided an average speedup of 14$\times$ over static HLS and 4$\times$ over dynamic HLS.


\begin{acks}
This work was partly supported by a PhD scholarship from the UK EPSRC. 
We thank the anonymous reviewers and our shepherd Louis-Noël Pouchet for improving this paper.
We thank Intel for free access to FPGAs through their FPGA DevCloud.

\end{acks}

\bibliographystyle{ACM-Reference-Format}
\balance
\bibliography{references}


\end{document}